\documentclass[a4paper]{jpconf}
\usepackage{graphicx}
\begin{document}
\title{A parameterisation of single and multiple muons in the deep water or ice}
\author{A Margiotta}
\address{Dipartimento di Fisica dell'Universita and Sezione INFN di Bologna, \\ viale C. Berti-Pichat, 6/2, 40127 Bologna, Italy}
\ead{Annarita.Margiotta@bo.infn.it}
\begin{abstract}
A new parameterisation of atmospheric muons deep underwater (or ice) is presented. It takes into account the simultaneous arrival of muons in bundle giving the  multiplicity of the events and the muon energy spectrum as a function of their lateral distribution in a shower.
\end{abstract}
\section{Introduction}
Primary cosmic rays interacting with atmospheric nuclei produce electromagnetic and hadronic showers that, during their development through the atmosphere, originate a large number of neutral and charged particles. Although neutrino telescopes are located under large depths of water or ice, a great number of high energy atmospheric muons can reach the sensitive volume of the detectors \cite{antares}. Indeed, they are the most abundant signal in a neutrino telescope and can be used to calibrate the detector and to check its expected response to the passage of charged particles. On the other hand, they can represent  a dangerous background source for neutrino telescopes if track reconstruction programs, which use Cherenkov light emitted by downgoing muons, yield an upward going muon track \cite{icrc_ant}. \\
\indent A full Monte Carlo simulation, starting from the simulation of atmospheric showers, can accurately reproduce the main features of muons reaching a neutrino telescope, but requires a large amount of CPU time.  Some fast parameterisations of the atmospheric muon flux underwater are available in the literature \cite{okada,bugaev,klimu} but all of them do not take into account the simultaneous arrival of muons in bundles, which are particularly dangerous for misreconstructed tracks.\\
\indent In this work, parametric formulae are presented to evaluate the flux of atmospheric muons, taking into account the muon multiplicity and the muon energy spectrum in a bundle, as a function of the distance from the shower axis. A detailed description can be found in \cite{newpara}.
\section{The Monte Carlo simulation}
\label{sec:TheMonteCarloSimulation}
The  parameterisation of the multiple muon flux and energy spectra presented here relies on a full Monte Carlo simulation of primary Cosmic Ray (CR) interaction and shower propagation in the atmosphere (HEMAS code \cite{hemas}).
The adopted primary CR flux and composition model is described in \cite{horandel}.
The muons from the decay of secondary mesons reaching the sea level are then propagated to seven depths, from $2.0\ km \:w.e.$ down to $5.0\ km \:w.e.$, in steps of $0.5\ km \:w.e.$, using the MUSIC code \cite{music}.
The energy spectrum of muons depends on the vertical depth $h$, on the zenith angle $\theta$, on the muon multiplicity in the shower $m$ and on the distance of the muon from the shower axis $R$. 
Starting from the distributions of muon multiplicity, energy and distance of muons from the shower axis, a number of fits was performed in order to obtain the values of the constants to be used in the analytic description of the muon flux and energy.
\section{The parameterisation}
\begin{itemize}
	\item 
The multiplicity distribution is reproduced, as a function of water depth and zenith angle, following an analytic description used by the Frejus collaboration \cite{frejus}:
\begin{equation}
\Phi(m;h,\theta)= {K(h,\theta) \over m^{\nu(h,\theta)}} \quad \rm{with} \quad \nu=  {\nu_1 \over (1+\Lambda \cdot m)} 
\label{eq:eq1}
\end{equation}
$\Lambda \simeq 0$ is compatible with all our fits.
9 constants are necessary to describe eq. \ref{eq:eq1} in terms of h and $\vartheta$. From (\ref{eq:eq1}), the  depth-intensity relation for vertical muons ($\vartheta =0^o$) can be calculated. A comparison with other parameterisations is shown in fig. 1. In fig. 2  comparisons of the zenith distribution evaluated with eq. \ref{eq:eq1} and some experimental data are shown.
\begin{figure}[h]
\begin{minipage}{18pc}
\includegraphics[width=18pc]{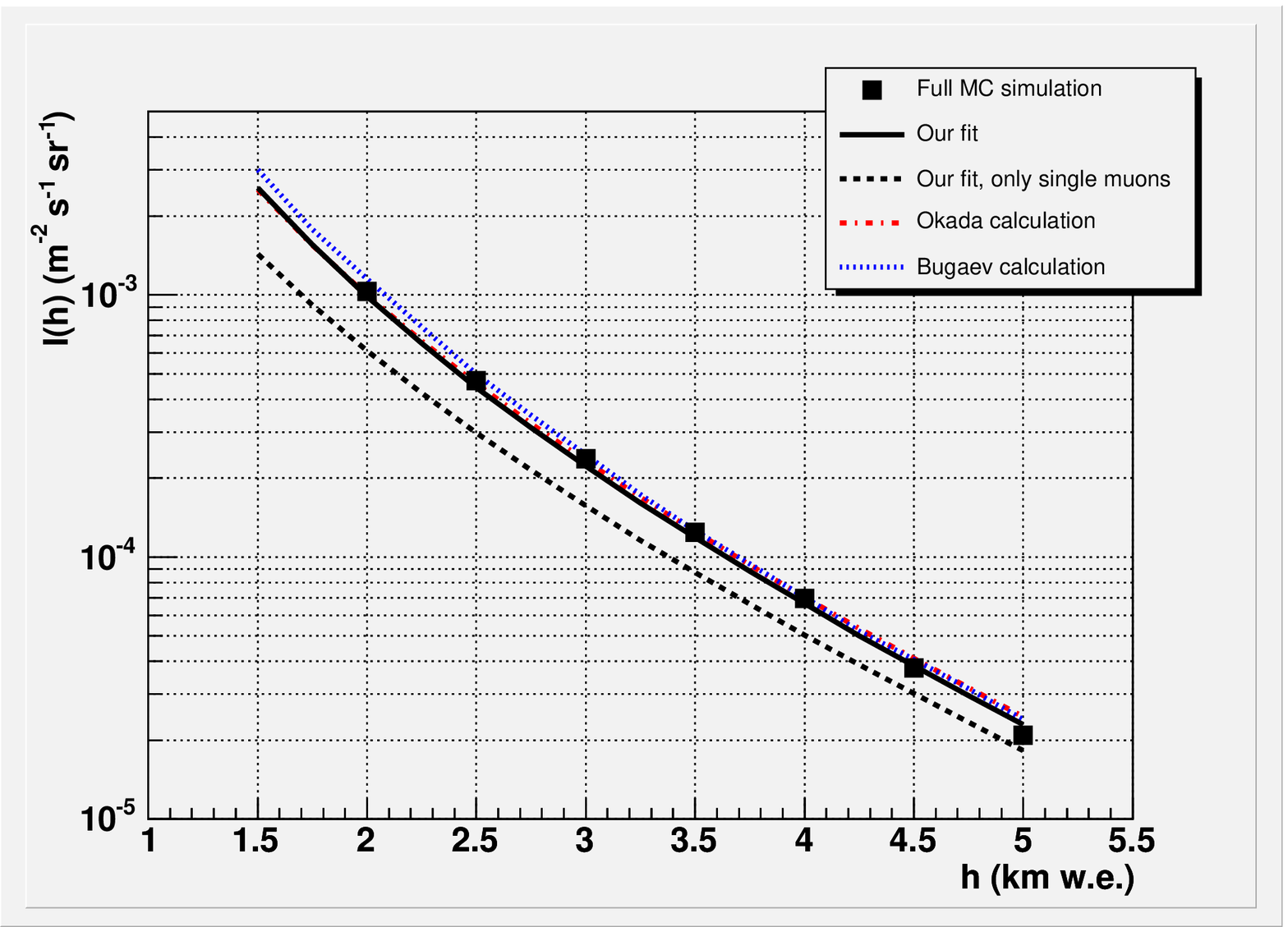}
\caption{\label{flux1}Muon vertical intensity versus  depth. 
Comparisons are shown among this parameterisation for single muon events (dashed line) and for the total muon flux (full line), other parameterisations: Okada \cite{okada} (dotted-dashed line), Bugaev et al. \cite{bugaev}(dotted line) and the full Monte Carlo simulation (points).}
\end{minipage}
\hspace{2pc}%
\begin{minipage}{18pc}
\includegraphics[width=18pc]{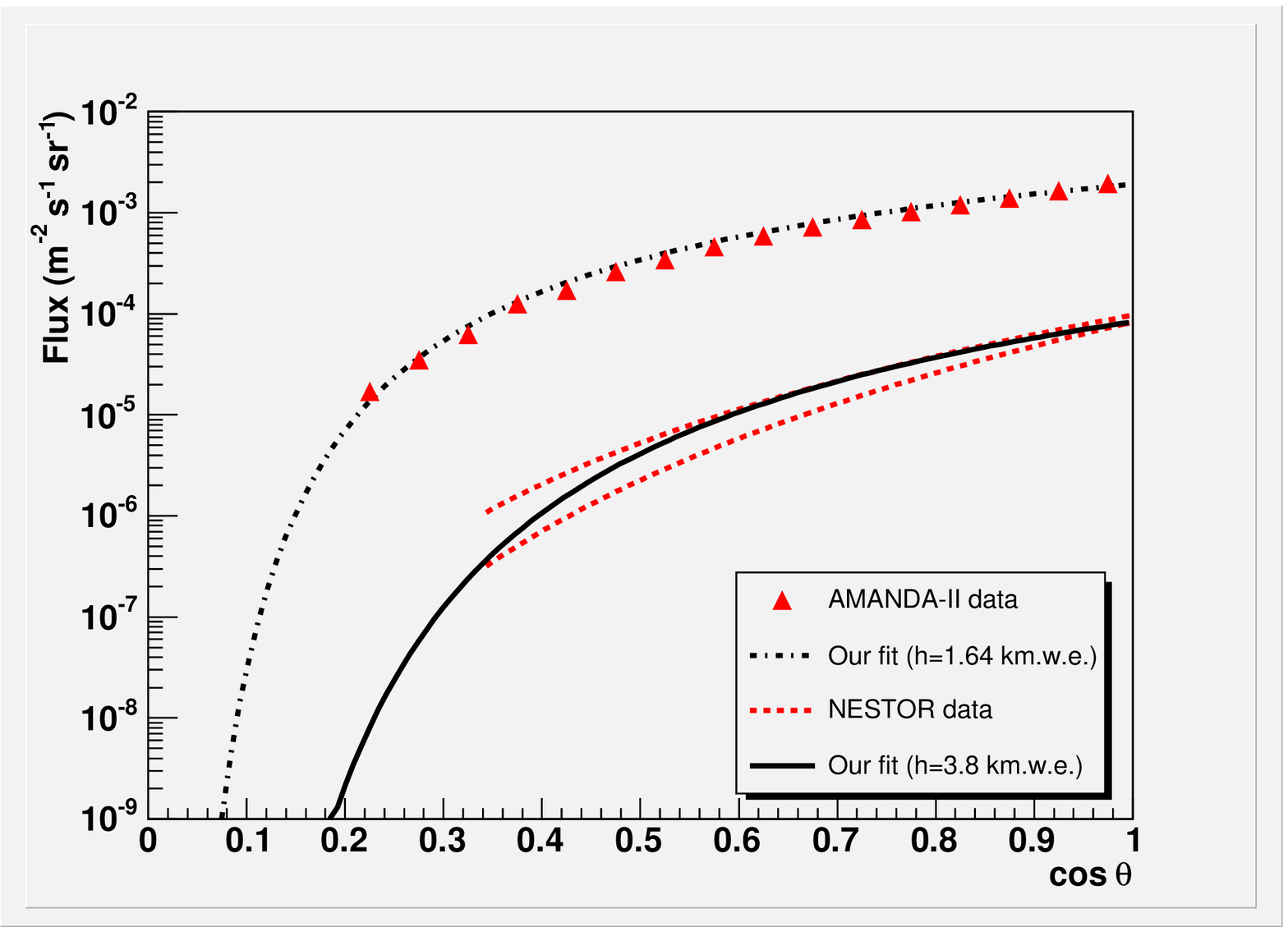}
\caption{\label{nestorama}Muon flux versus zenith angle. Dot-dashed line:   muon flux evaluated at  $h=1.64\ km \:w.e.$ compared with AMANDA-II measurements \cite{amanda} (triangles). Full line:  muon flux evaluated at $h=3.8\ km \:w.e.$ compared with the  angular distribution (1-sigma error) measured  by the NESTOR Collaboration \cite{nestor} (dashed lines).}
\end{minipage} 
\end{figure}
\item 
The energy spectrum of single muons is described by:
\begin{equation}
{ dN \over d (log_{10}E_\mu) } = G\cdot E_\mu e^{\beta X (1-\gamma)} [E_\mu + \epsilon (1-e^{-\beta X})]^{- \gamma}    
\label{eq:spectrum} 
\end{equation}
(\cite{lista}), where $\gamma$ is the spectral index of the primary beam and $\epsilon= \alpha /\beta$ is defined from $-\langle{dE(E_\mu) \over dX}\rangle = \alpha + \beta E_\mu $. In this work, $\gamma$ and $\epsilon$ were used as fit parameters depending on h and $\vartheta$; $\beta$ was fixed to a value of 0.420 $(kmw.e.)^{-1}$.  6 constants are necessary in eq. \ref{eq:spectrum}.
\item 
The muon lateral spread in muon bundles is parameterised as 
\begin{equation}
{dN \over dR} = C { R\over (R+R_0)^\alpha }
\label{eq:radial} 
\end{equation}
$R_o$ and $\alpha$ can be expressed as functions of h, m and $\vartheta$ using 9 constants. The description of the muon lateral spread is the preliminary step to evaluate the muon energy distribution in a bundle.
\item The energy spectrum of muons arriving in bundles has the same general form as for single muons, but each parameter shows a dependence on multiplicity and distance from the axis and not only on depth and zenith angle. 15 constants are necessary in this case.
\end{itemize}

A comparison between the average energy of single and double muons evaluated with this parameterisation and the results of MACRO measurements, performed with a Transition Radiation Detector \cite{macrotrd2} under different depths of rock, is presented in Tab. 1. A correction has been applied to take into account the differences in energy losses of muons in standard rock and water.  
\begin{table}[h]
\begin{center}
\begin{tabular}
{|c|c|c||c|c||c|c|} \hline
    &                  &             &\multicolumn{2}{c||}{Single muons}&\multicolumn{2}{c|}{Double muons} \\ \hline
$h $&$\overline \theta$&$\overline R$&$\overline E^{MACRO}_{1\mu,Rock}$ & $\overline E_{1\mu,W}$(corr.)& $\overline E^{MACRO}_{2\mu,Rock}$ & $\overline E_{2\mu,W}$(corr.) \\
$(hg\ cm^{-2})$  & (deg) & $(m)$ &  $(GeV)$ & $(GeV)$& $(GeV)$& $(GeV)$\\ \hline
3280 & 3 & 5.2 &  $250\pm 17$ & 248 & $321\pm 23$ & 329 \\
3420 & 10& 4.7 &  $262\pm 18$ & 254 & $366\pm 24$ & 350 \\
3600 & 20& 4.3 &  $278\pm 19$ & 265 & $400\pm 25$ & 383 \\
3800 & 30& 4.0 &  $283\pm 19$ & 278 & $417\pm 25$ & 412 \\ \hline 
\hline
\end{tabular}
\end {center}
\caption{\small Average energy of single (column 4) and double (column 6) muons measured by the MACRO collaboration \cite{macrotrd2} for different values of the slant depth (column 1) and of the average muon zenith angle direction (column 2). The predictions of this work   are given in columns 5 and 7 for $m=1$ and $m=2$, respectively. The calculated values have been corrected to take into account the muon energy loss differences in water and rock. For double muons, the computed average distance of muons from the shower axis, reported in column 3, has been assumed.}
\label{tab:trd}
\end{table}
\section{Conclusions}
A new parameterisation of the atmospheric muon flux under water, considering the simultaneous arrival of muons in bundles, is presented.   Energy and zenith distributions  obtained with these formulae are in good agreement with the predictions from other parameterisations, with the full Monte Carlo and with  experimental measurements. This parameterisation can be used as a fast generator to simulate atmospheric muon fluxes in underwater/ice neutrino telescopes.

\end{document}